\documentclass[conference]{IEEEtran}
\usepackage{cite}

\ifCLASSINFOpdf
\else
\fi
\usepackage{amsmath}
\usepackage{algorithmic}

\usepackage{array}
\usepackage{url}

\usepackage{enumitem}
\usepackage{amsfonts}
\usepackage{pifont}
\usepackage{algorithm}
\usepackage{graphicx}
\usepackage{graphics}
\usepackage{multirow}
\usepackage{balance}
\usepackage{hyperref}
\usepackage{booktabs}

\begin{document}
%
\title{LUT-LLM: Efficient Language Model Inference with Memory-based Computations on FPGAs}

\author{
\IEEEauthorblockN{
Zifan He\IEEEauthorrefmark{1},
Shengyu Ye\IEEEauthorrefmark{2},
Rui Ma\IEEEauthorrefmark{2},
Yang Wang\IEEEauthorrefmark{2} and
Jason Cong\IEEEauthorrefmark{1}}

\IEEEauthorblockA{\IEEEauthorrefmark{1}University of California, Los Angeles}
\IEEEauthorblockA{\IEEEauthorrefmark{2}Microsoft Research\\
Email: zifanhe1202@g.ucla.edu, \{v-yeshengyu, mrui, Yang.Wang92\}@microsoft.com, cong@cs.ucla.edu}
}

\maketitle

\begin{abstract}
  The rapid development of large language models (LLM) has greatly enhanced everyday applications. While many FPGA-based accelerators, with flexibility for fine-grained data control, exhibit superior speed and energy efficiency compared to GPUs, recent GPU-specific optimizations have diminished this advantage. When limited to arithmetic-based computation, FPGAs often underperform GPUs due to their comparatively fewer computational resources. To address this challenge, we exploit a key advantage of FPGAs over GPUs: abundant distributed on-chip memory embedded among computational units. We believe that shifting LLM inference from arithmetic-based to memory-based computations through table lookups can improve the efficiency on FPGAs to compete with GPUs. However, existing methods are inefficient or unable to scale and deploy language models due to algorithm and architecture design limitations. This paper introduces \textbf{LUT-LLM}\footnote{Github: \url{https://github.com/LUT-FPGA/LUT-LLM}}, the first FPGA accelerator that deploy 1B+ language model with memory-based computation, leveraging vector quantization. We construct a performance model, evaluate multiple quantization schemes, and identify activation-weight vector co-quantization as the most effective approach. To support this scheme, LUT-LLM features (1) bandwidth-aware parallel centroid search to reduce decoding latency, (2) efficient 2D table lookups, and (3) a spatial-temporal hybrid design to reduce data caching for a higher throughput table lookup. We develop a training recipe that converts existing models to support table lookups with high accuracy and prototype LUT-LLM for Qwen 3 1.7B model on the AMD V80 FPGA, reducing arithmetic operations by $4\times$ and achieving a $1.10\sim3.29\times$ faster generation speed and a $3.05\sim 6.60\times$ higher energy efficiency than GPUs. 
\end{abstract}


%
\IEEEpeerreviewmaketitle

\section{Introduction}

The advancement of LLMs has enabled a wide range of applications such as conversational chatbots \cite{dam2024complete, oruche2025survey}, coding agents \cite{zhang2024codeagent, li2024codetree}, and deep research \cite{huang2025deep} that assist our daily lives. Recently, FPGA have been widely recognized as promising platforms for efficient single-batch LLM inference compared with GPUs \cite{chen2024understanding, chen2024allo, zeng2024flightllm}. This covers edge scenarios such as personal AI assistants \cite{cheng2024autopal, lee2024towards}, smart home devices \cite{qualcomm2025smarthomes}, and interactive robotics \cite{ayub2024interactive}. Nevertheless, the advancements in GPU kernel and algorithm optimization have narrowed this performance gap. In particular, serving a Llama 2 7B model on a single NVIDIA A100 GPU obtains a $2.38 \times$ throughput improvement and a $3.27 \times$ energy efficiency gain when FlashAttention \cite{dao2023flashattention}, FlashDecoding \cite{hong2023flashdecoding++}, and quantization with GPTQ \cite{frantar2022gptq} are enabled. Under these settings, GPU inference is $2 \times \sim 6.37 \times$ faster with similar or better energy efficiency than state-of-the-art FPGA accelerators \cite{chen2024allo, zeng2024flightllm}. These results highlight the growing competitiveness of GPUs in efficient inference and suggest that sustaining FPGA's advantages will require algorithmic and architectural innovations.

\begin{figure}[t]
    \centering
    \includegraphics[width=0.9\linewidth]{ 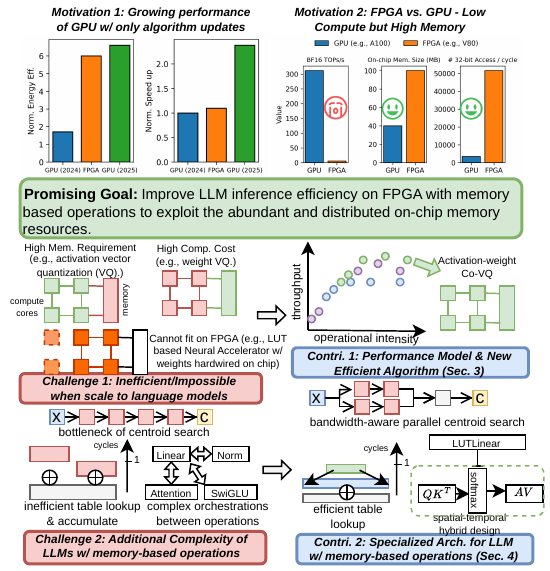}
    \caption{Motivations and challenges of memory-based computation for LLM inference on FPGA, with the corresponding solutions as the technical contributions in LUT-LLM.}
    \label{fig:challenge}
\end{figure}

When competing with GPUs, the major challenge is that FPGAs often have fewer computational resources than GPU \cite{amd2023alveou280, nvidia2020v100, amd2024alveov80, nvidia2021a100}. This leads to slow input processing and underutilized memory bandwidth when FPGAs spend more time on computation than on memory access. However, FPGAs typically provide larger amounts of distributed on-chip memory units: the AMD V80 (7nm) \cite{amd2024alveov80} integrates $14.9\times$ more on-chip memory units with a $2.5\times$ larger capacity than the NVIDIA A100 (7nm) \cite{nvidia2021a100}. Therefore, shifting the computational paradigm to \textbf{memory-based computations} is promising. Prior memory-based accelerators \cite{li2025lut, gerlinghoff2024table, andronic2025neuralut, miyasaka2024synthesis} show the efficiency on neural networks within 10M parameters, but face two challenges when deploying language models (Figure \ref{fig:challenge}):

\begin{itemize}[wide=0pt]
    \item \textbf{Scaling inefficiency and infeasibility (C1)}: direct deployment of existing methods that convert models for table lookups \cite{li2025lut, liu2024vptq, van2024gptvq} can lead to inefficiency due to higher memory and computation cost compared to conventional arithmetic approaches, or be infeasible with limited on-chip resources.
    \item \textbf{Additional Complexity of LLMs (C2)}: prior memory-based neural accelerators overlook the additional complexity of LLM accelerations, where centroid search is hard to pipeline in decoding, lookup table access is limited by on-chip memory ports, and interactions between linear layers and other components further hinder data movement efficiency.

\end{itemize}

In this paper, we propose \textbf{LUT-LLM}, the first FPGA accelerator targeting 1B+ parameter LLMs using \textit{memory-based computation}. LUT-LLM replaces conventional linear layers with table lookups over pre-computed dot-product results. To enable efficient table construction, we adopt \textbf{vector quantization}~\cite{gray1984vector}, which maps multiple values to low-bit indices. This design \textbf{reduces the number of operations} and \textbf{shortens per-operation latency}, enabling competitive performance compared to GPUs that often underutilize available memory bandwidth~\cite{zeng2024flightllm}. To address \textbf{C1}, we develop a performance model for LLM inference with memory-based computation and vector quantization to understand the scaling (Section~\ref{sec:performance_model}). We innovatively show that vector quantizing both weights and activations achieves the best performance while maintaining comparable accuracy. From this foundation, we design the LUT-LLM accelerator with three features that resolve \textbf{C2} (Section \ref{sec:arch}):

\begin{itemize}[wide=0pt]
  \item \textbf{Bandwidth-aware parallel centroid search:} LUT-LLM finds the closest centroid of every input vector to access the corresponding entry in the lookup tables. Unlike a naive single pipeline \cite{li2025lut} or a complete binary reduction tree architecture, LUT-LLM employs a hybrid architecture with multiple parallel pipelines and a smaller reduction tree. The parallelization size and pipeline depth are co-designed with the memory bandwidth to maximize throughput while hiding search latency.

  \item \textbf{Efficient 2D table lookup based prefix-sum:} To support vector quantization of both activations and weights, we construct 2D lookup tables, where one dimension corresponds to activation centroid indices and the other to weight centroid indices. During inference, the tables are loaded to on-chip memory in parallel with the centroid search, and the corresponding table entries are dynamically retrieved and expanded online, along with SIMD accumulation.

  \item \textbf{Spatial-temporal hybrid design:} LUT-LLM adopts a hybrid execution strategy, following a design principle similar to InTAR \cite{he2025intar} and SSR \cite{zhuang2024ssr}. Specifically, the attention layer is implemented in a dataflow manner, whereas linear layers are executed sequentially. This design spares $14\%$ on-chip buffers from attention to the linear layer for more parallel accesses.

\end{itemize}
We develop a training recipe for vector-quantized model for our scheme to convert and deploy Qwen 3 1.7B on LUT-LLM prototyped with AMD V80 FPGA. Compared with GPUs at the same technology node, our design achieves a $1.10-3.29 \times$ speedup and $3.05-6.60\times$ higher energy efficiency than AMD MI210 and NVIDIA A100.


\section{Background}

\subsection{Language Model Acceleration}

Most LLM accelerators adopt the transformer architecture \cite{vaswani2017attention}, which contains a series of attention blocks implemented in grouped-query attention (GQA) \cite{ainslie2023gqa} and feedforward networks (FFN) with residual connections \cite{chen2024understanding}. Previous works \cite{chen2024allo, zeng2024flightllm, he2025intar, hong2022dfx} develop customized transformer accelerators with arithmetic-based operations. FlightLLM and DFX \cite{hong2022dfx, zeng2024flightllm} sequentially execute operations with various sparsification and quantization techniques. Allo \cite{chen2024allo} and StreamTensor \cite{ye2025streamtensor} are dataflow accelerator design frameworks that can be used for LLMs, and InTAR \cite{he2025intar} proposes a spatial-temporal hybrid design for efficient on-chip management of LLMs.

\subsection{Weight Vector Quantization}

\begin{figure}[ht]
    \centering
    \includegraphics[width=0.9\linewidth]{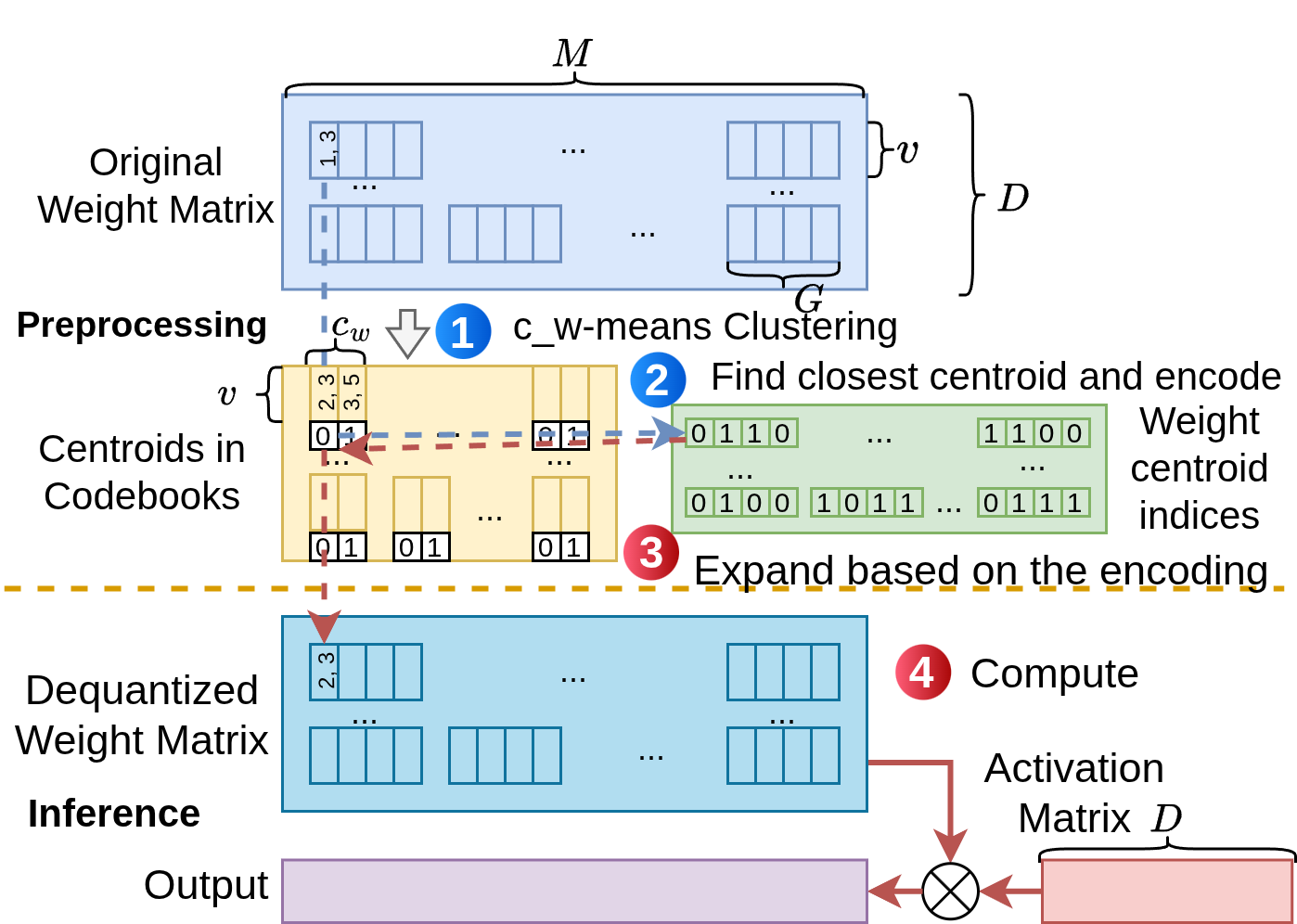}
    \caption{Linear projection with weight vector quantization.}
    \label{fig:vq_weight}
\end{figure}

Conventional quantization compresses scalars into low-bit representations via scaling and shifting \cite{nagel2020up, lin2024awq, xiao2023smoothquant, frantar2022gptq}. In contrast, vector quantization (VQ) \cite{gray1984vector, abouelhamayed2024pqa} encodes groups of $v$ elements as indices to representative centroids. As seen in Figure \ref{fig:vq_weight}, in the preprocessing stage, the quantizer \ding{192} partitions the matrix into length-$v$ vectors, clusters them into $c_w$ centroids forming a \textbf{codebook}, and \ding{193} replaces each vector with the index of its closest centroid. The resulting \textbf{weight centroid index table} stores indices at a much lower bitwidth. For example, with $v=2$, vector $(1,3)$ is encoded as index $0$, mapped to centroid $(2,3)$.

For language models, VQ is applied to weight matrices to reduce memory cost \cite{van2024gptvq, tseng2024quip, tseng2024qtip, liu2024vptq}. An $M \times D$ matrix is divided into groups of $G$ vectors, each with its own codebook and index table (e.g., $G=4$ in Figure \ref{fig:vq_weight}). In inference, the device \ding{194} loads codebooks and indices, reconstructs the weights by centroid lookup, and \ding{195} multiplies the recovered matrix with activations. Compared to scalar quantization, VQ provides non-uniform compression \cite{van2024gptvq}, yielding higher representational power for superior accuracy at the same bitwidth.

\subsection{Memory-based Computation}

Memory-based computation replaces arithmetic by loading pre-computed outputs from memory, making it well-suited for on-chip memory-rich platforms such as FPGAs. It reduces resource usage and latency since a multiply-accumulate (MAC) operation (two reads, one multiply and one addition) can be replaced by a single memory read. This also lowers energy: a memory-based MAC consumes only 3.8 pJ in 7 nm, $2.4\times$ less than its arithmetic counterpart \cite{jouppi2021ten}. Moreover, memory-based LLMs avoid online dequantization because values are pre-computed in full precision. The pre-computed lookup table can be further quantized with low loss \cite{li2025lut}, used in our design. 

\begin{figure}[ht]
    \centering
    \includegraphics[width=\linewidth]{  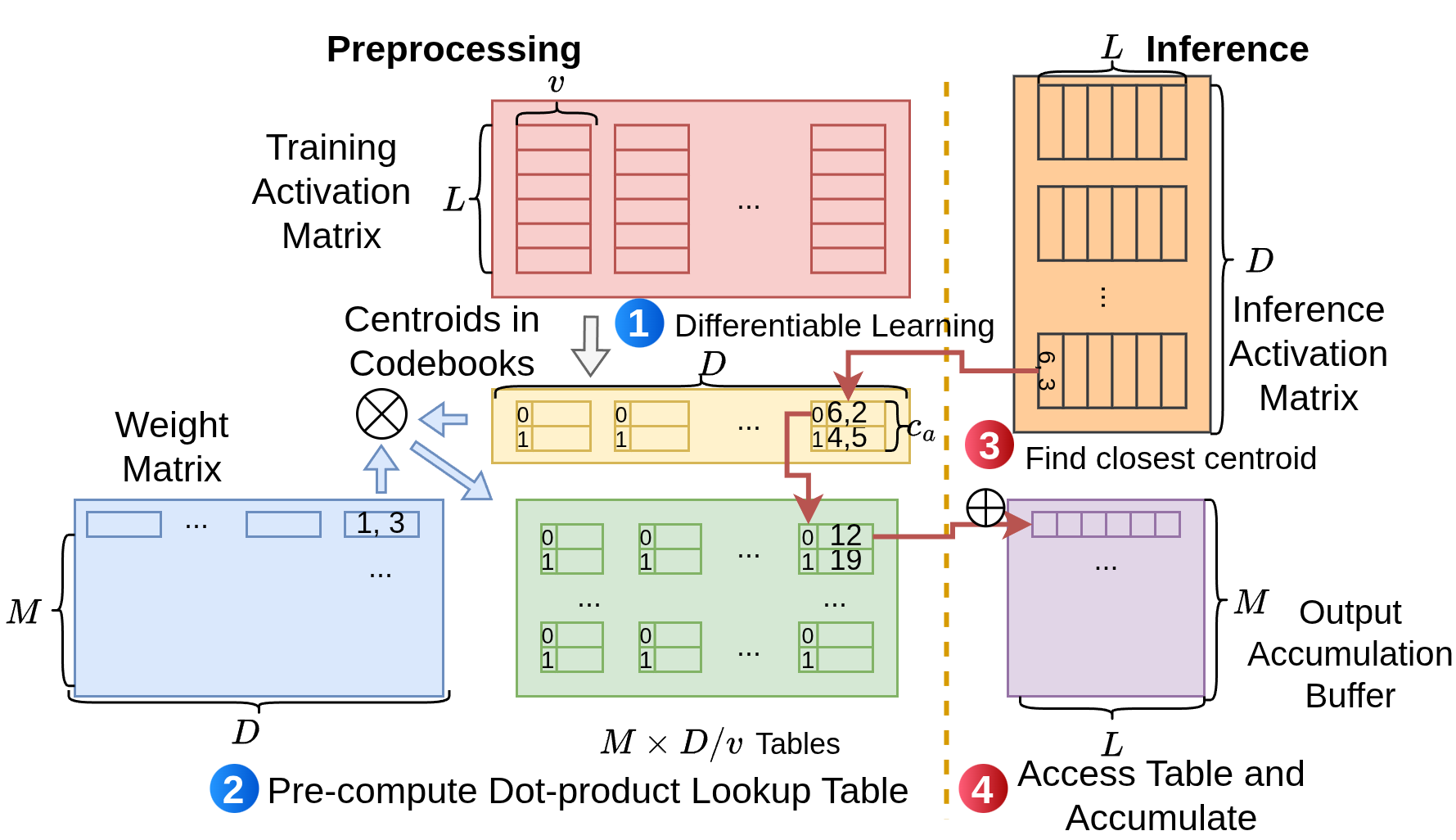}
    \caption{Linear projection with activation VQ. Precomputed dot products between weight matrix and centroids are stored in the lookup tables.}
    \label{fig:vq_act}
\end{figure}

Previous works, including TLMAC \cite{gerlinghoff2024table} and NeuraLUT \cite{andronic2025neuralut}, consider using memory-based computations for simple neurons with FPGA soft logic. However, the limited resource makes it infeasible to scale to language models that compete with GPUs. For LLMs, weight quantization in these methods requires large storage for offline lookup tables due to the wide range of activations. Recent works \cite{li2025lut, tang2023lut, wei2025t, raha2017qlut} therefore apply VQ to activations in linear layers. As illustrated in Figure~\ref{fig:vq_act}, the quantizer collects activation samples during training. To preserve accuracy, methods such as LUT-NN \cite{tang2023lut} \ding{192} learn codebooks with $c_a$ centroids through differentiable training and \ding{193} pre-compute dot products between weight vectors and centroids, storing them as lookup tables. For example, a weight vector $(1,3)$ multiplied by centroids $(6,2)$ and $(4,5)$ yields entries $12$ and $19$. During inference, the quantizer \ding{194} identifies the nearest centroid for each input vector, \ding{195} retrieves the corresponding table entry, and accumulates the results. For instances, input $(6,3)$ maps to centroid $(6,2)$ at index $0$, retrieving $12$. LUT-DLA \cite{li2025lut} is a hardware-algorithm co-design that improves model learning and accelerator design. Though effective in CNNs, it still requires non-trivial large storage for LLMs, which limits decode and short-context prefill performance. In addition, its single-pipeline centroid search creates bottlenecks in the decode stage. 

\section{Algorithm and Performance Modeling} \label{sec:performance_model}

In this section, we illustrate the performance model of our new quantization scheme to guide our hardware design in Section \ref{sec:arch}. Detailed derivation is omitted to save space.

\subsection{Performance of Vector Quantized Linear Layer}

We start with modeling the performance of the linear layer, which is affected by the vector quantization, and then extend this to the complete language model. Table \ref{tab:symbols} illustrates the symbols utilized in the performance model. The equivalent quantization scheme is $\log(c_w)/v$-bit for weight and $\log(c_a)/v$-bit for activation. Here, we mainly show the key results for different quantization strategies, and leave the detailed derivation in Appendix \ref{sec:derivation}.

\begin{table}[ht]
\centering
\caption{Symbols used in performance modeling of linear layer}
\label{tab:symbols}
\resizebox{\linewidth}{!}{
\begin{tabular}{ll|ll}
\hline
\textbf{Symbol} & \textbf{Meaning} & \textbf{Symbol} & \textbf{Meaning} \\
\hline
\multicolumn{2}{c|}{\textit{Used Hardware Resource}} & \multicolumn{2}{c}{\textit{Quantization Configurations}} \\
$N_{p}$     & On-chip memory ports & $G$       & Vectors per quantization group \\
$b_{p}$     & Bit-width per access  & $v$       & Vector Length \\
$N_{c}$     & Compute units & $c_{w}$ & Centroids per weight codebook \\
$Op_{fp32}$ & FP32 MACs/cycle/compute unit   & $c_{a}$   & Centroids per activation codebook \\
$Op_{int8}$ & INT8 MACs/cycle/compute unit  & $D, M$ & Input/Output weight dimensions \\
$C$         & Off-chip bandwidth (bytes/cycle) & $L$ & Activation (Sequence) length \\
\hline
\end{tabular}
}
\end{table}

For \textbf{weight vector quantization}, the latency of loading codebook and weight centroid indices is
\begin{equation} \label{eq:w_mem}
    T_{mem} = (4MDc_w/Gv + MD\log(c_w)/8v) / C
\end{equation}
and the latency of on-chip computation is 
\begin{equation} \label{eq:w_lat}
    \begin{split}
        T_{lat} &= \frac{MD(\frac{\log(c_w)}{v}+\frac{32}{Gv})}{N_{p}b_{p}}+\frac{MDL}{\min(N_{c}Op_{fp32}, \frac{N_{p}b_{p}}{32})}
    \end{split}
\end{equation}
When overlapping off-chip memory access with on-chip computations by double buffering, the overall latency is $\max(T_{mem}, T_{lat})$. For example, with $M = 512$, $D = 32$, $G = 256$, $c_w = 16$, $v = 2$, $L = 1$, $C = 64$, and when we assign 16 memory ports of 32-bit wide memory and 256 FP32 compute units, $T_{mem} = 66$ and $T_{lat} = 1090$, resulting in $1090$ cycles for overall latency.

For \textbf{activation vector quantization}, the latency of loading codebook and lookup tables is
\begin{equation} \label{eq:a_mem}
    T_{mem} = (MDc_a/v + 4Dc_a/v)/C
\end{equation}
When having $S$ parallel centroid searches, the latency of table lookup per $S$ parallel centroid search and accumulation is:
\begin{equation} \label{eq:a_tl}
    \begin{split}
    T_{tl} &= SML/\min(SM, N_{p}b_{p}/8) \\
        &+ \frac{SML}{\min(SM, (N_{c} - Sc_av/Op_{fp32})Op_{int8}, N_pb_p/8)}
    \end{split}
\end{equation}
The total latency of on-chip computation is
\begin{equation} \label{eq:a_lat}
    T_{lat} = \min_{S}(D \max(\log(c_a)+L-1, T_{tl})/S)
\end{equation}
With the same setting as the previous example and $c_a = 64$, $T_{mem} = 8256$ and $T_{lat} = 512$, leading to $8256$ cycles for overall latency. Notice that some table-lookup-based FPGA accelerators \cite{gerlinghoff2024table}, while quantizing weights, can be modeled in this formula by replacing $c_a/v$ with $2^b$ and divide $D$ by quantization group size, where $b$ is the activation bitwidth.

\begin{figure}[ht]
    \centering
    \includegraphics[width=\linewidth]{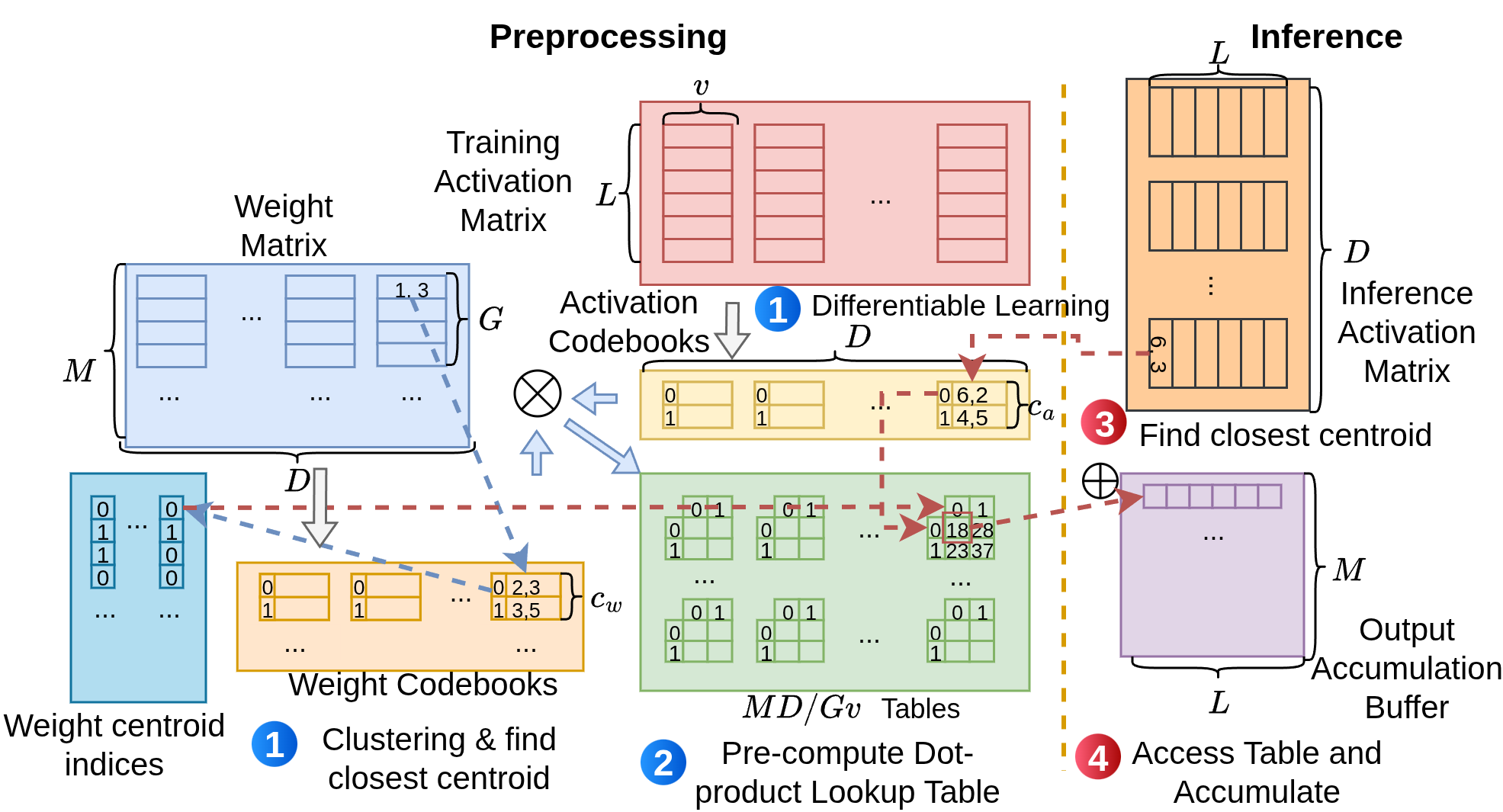}
    \caption{Linear projection with both activation and weight vector quantization with 2D lookup tables.}
    \label{fig:vq2d}
\end{figure}

From the analysis in this example, we observe that either weight-only or activation-only VQ applied in prior works \cite{van2024gptvq, tang2023lut} incur high computation latency or require loading large tables. To resolve these issues, we extend these schemes to \textbf{activation-weight co-quantization}, which quantizes in coarser-grain but still maintain comparable accuracy as shown in Table \ref{tab:algo_results}. As illustrated in Figure \ref{fig:vq2d}, the process involves \ding{192} generating codebooks for weights and activations, \ding{193} pre-computing dot products between them to form 2D lookup tables, \ding{194} finding the closest centroid for each activation vector, and \ding{195} accessing the entry based on the activation and weight centroid indices for accumulation. In the figure, vector $(6,3)$ is closer to $(6,2)$ at index 0. As the first weight centroid index is 0, we access the lookup table entry $(0,0)$ with value $18$.

With this scheme, the latency of off-chip memory access is 
\begin{equation} \label{eq:2d_mem}
    T_{mem} = (MD\tfrac{c_ac_w}{Gv} + MD\log(c_w)/8v + 4Dc_a/v)/C
\end{equation}
and the on-chip computation latency is
\begin{equation}
    \begin{split}
    T_{tl} &= SML/G\min(SM/G, N_{p}b_{p}/8) \\
        &+ \frac{SML}{\min(SM, (N_{c} - Sc_av/Op_{fp32})Op_{int8}, N_pb_p/8)}
    \end{split}
\end{equation}
\begin{equation} \label{eq:2d_lat}
    T_{lat} = \min_{S}(\tfrac{D}{S} \max(\log(c_a)+L-1, T_{tl}))
\end{equation}
The same setting as previous example results $T_{mem} = 569$, $T_{lat} = 288$, and the overall latency is $569$ cycles.

\subsection{Analysis and Comparison}

\begin{figure}
    \centering
    \includegraphics[width=0.95\linewidth]{  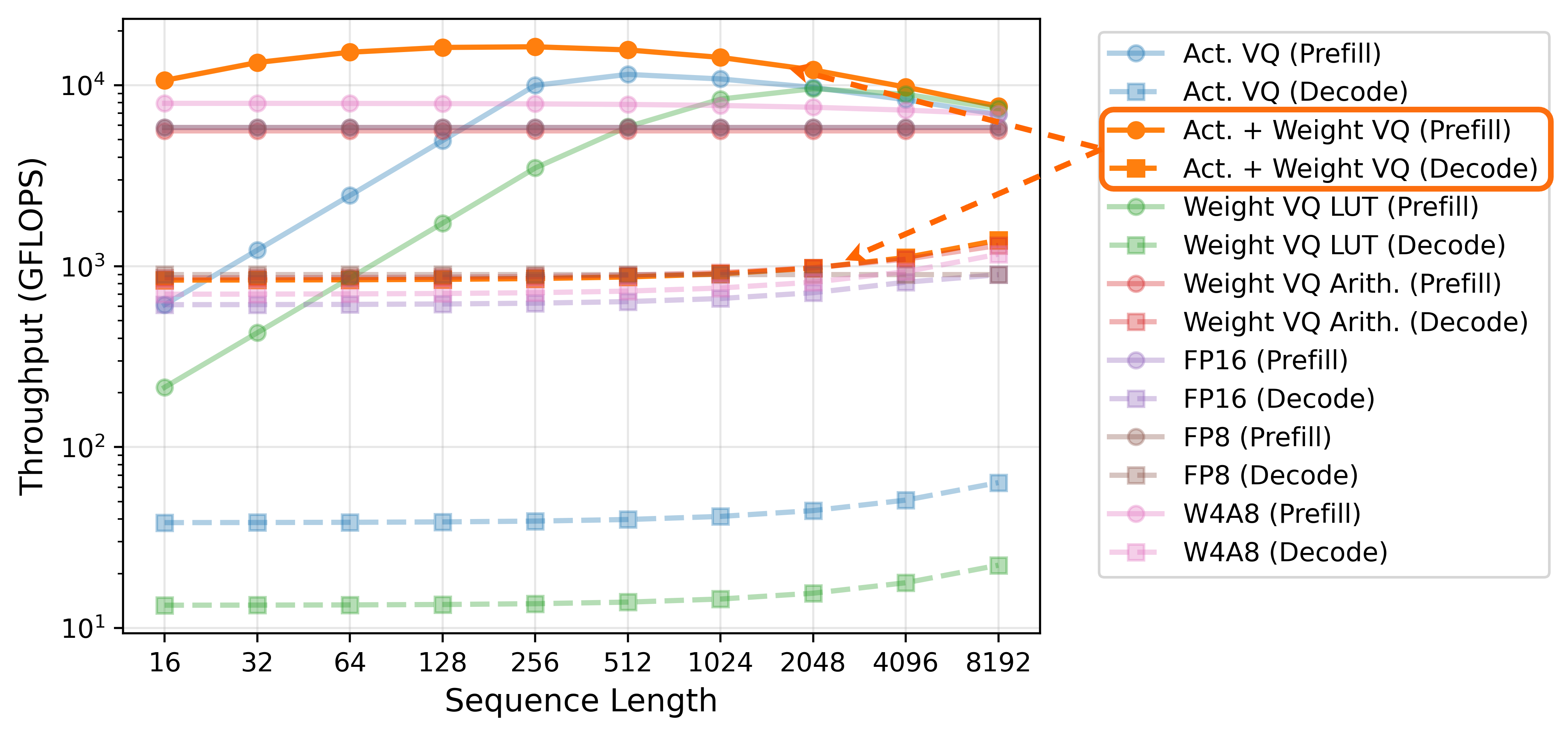}
    \caption{Normalized throughputs of the prefill and decode stage for Qwen 3 1.7B on AMD V80 FPGA with various quantization schemes. Activation-weight VQ can achieve superior performance than other schemes in both prefill and decode. FP8/FP16 are not natively supported by DSPs on V80 and require online conversion to FP32 or custom compute logics \cite{liu2025flightvgm, zeng2024flightllm, hong2022dfx}.}
    \label{fig:comparison}
\end{figure}

We extend the performance model of vector-quantized linear layers to a complete transformer architecture following \cite{chen2024understanding}, with additional modeling for newly introduced computation blocks in modern LLMs (e.g., GQA, SwiGLU, etc.). Following prior work \cite{wei2025t, wang2023bitnet, wang2024bitnet, lin2024qserve}, attentions and non-linear operations require floating point for accuracy. Figure \ref{fig:comparison} depicts the roofline analysis of deploying the Qwen 3 1.7B model on the AMD V80 FPGA with a naive memory-based computation (Act. VQ) like \cite{li2025lut, wei2025t} and arithmetic-based computation (FP16). The quantization configuration is: $c_a=64$, $v=2$, INT8 lookup tables, and FP32 codebooks. Compared with conventional roofline analyzes of transformer acceleration \cite{williams2009roofline, chen2024understanding} with arithmetic-based computations, we observe two properties of the naive memory-based computation for the vector quantized model. First, memory-based compute can have higher throughput than arithmetic-based compute throughput with a longer sequence in the prefill stage, leading to at most $1.7\times$ performance boost compared to arithmetic-based compute. This is because memory-based compute utilizes memory access to compute dot products, increasing the total available compute resources. The throughput converges to the compute bound as the sequence length increases, since the computation is dominated by attention layers. Second, naive memory-based computing needs to load lookup tables and codebooks that are $\mathbf{16\times}$ larger than the weights of FP16. This results in significantly \textbf{lower operational intensity and throughput} in decode and short-context prefill. Such behavior is unfavorable for tasks involving short inputs or long outputs (e.g., short question-answering \cite{rajpurkar2016squad} and creative writing \cite{creative-writing-bench-v3}).


We then investigate other quantization schemes for a more efficient alternative on FPGA such as W4A8, FP8, Weight VQ, and Activation + Weight VQ. Figure \ref{fig:comparison} compares the normalized throughput of these vector quantization methods with conventional scalar quantization on FPGA accelerators \cite{hong2022dfx, he2025intar, chen2024allo}, using $G=512$ and $c_w=16$ for weight VQ. Weight VQ is employed with either arithmetic operations used in most GPU implementation as illustrated in Figure \ref{fig:vq_weight}, or with memory-based operations by table lookups \cite{gerlinghoff2024table}. Weight VQ with arithmetic operations boosts decoding throughput by reducing off-chip memory access, but its prefill performance remains constrained by costly floating-point operations, and decoding is further limited by on-chip memory bandwidth. Weight VQ with table lookups, while mitigating these issues, requires large table storage, similar to activation VQ. In contrast, \textbf{activation-weight co-quantization} realizes the highest throughput across both stages than other scalar and vector quantization schemes. This improvement arises from three factors: (1) multiple weight vectors map to the same centroid vector, reducing the total number of entries required in the lookup tables and easing the off-chip memory bandwidth demand; (2) co-quantization has higher data reuse of lookup tables than weight-only or activation-only quantization due to repeating access to the same centroid, lowering memory port requirements to sustain high parallelism; and (3) INT8 table lookup and accumulation reduce on-chip memory pressure and eliminate FP dequantization operations requried by W4A8 scheme, ensuring high computational throughput on FPGAs.



\section{LUT-LLM Architecture} \label{sec:arch}

\subsection{Architecture Overview}

Based on the performance analysis, we design the LUT-LLM accelerator with activation-weight co-quantization applied. Since not all operations can be executed through table lookup and the language model involves mixed precisions and computation orders, we carefully orchestrate the execution of operations and allocate on-chip resources to ensure its feasibility on the target FPGA. LUT-LLM integrates a LUTLinear engine with a global on-chip buffer, a dataflow attention engine, and two special function units (SFUs) for non-linear operations (SwiGLU and LayerNorm). The LUTLinear engine performs all linear projections and routes the outputs. The dataflow attention engine employs two FP32 GEMM/GEMV engines to compute $QK^T$ and $\text{softmax}(QK^T)V$ in a pipelined manner, with rotational embedding and softmax applied in between. It supports both GEMM and GEMV to sustain high throughput in both the prefill and decode stages within a single design. The SFUs handle nonlinear operations to process the output embeddings from the LUTLinear engine. 

\begin{figure}[t]
    \centering
    \includegraphics[width=0.8\linewidth]{  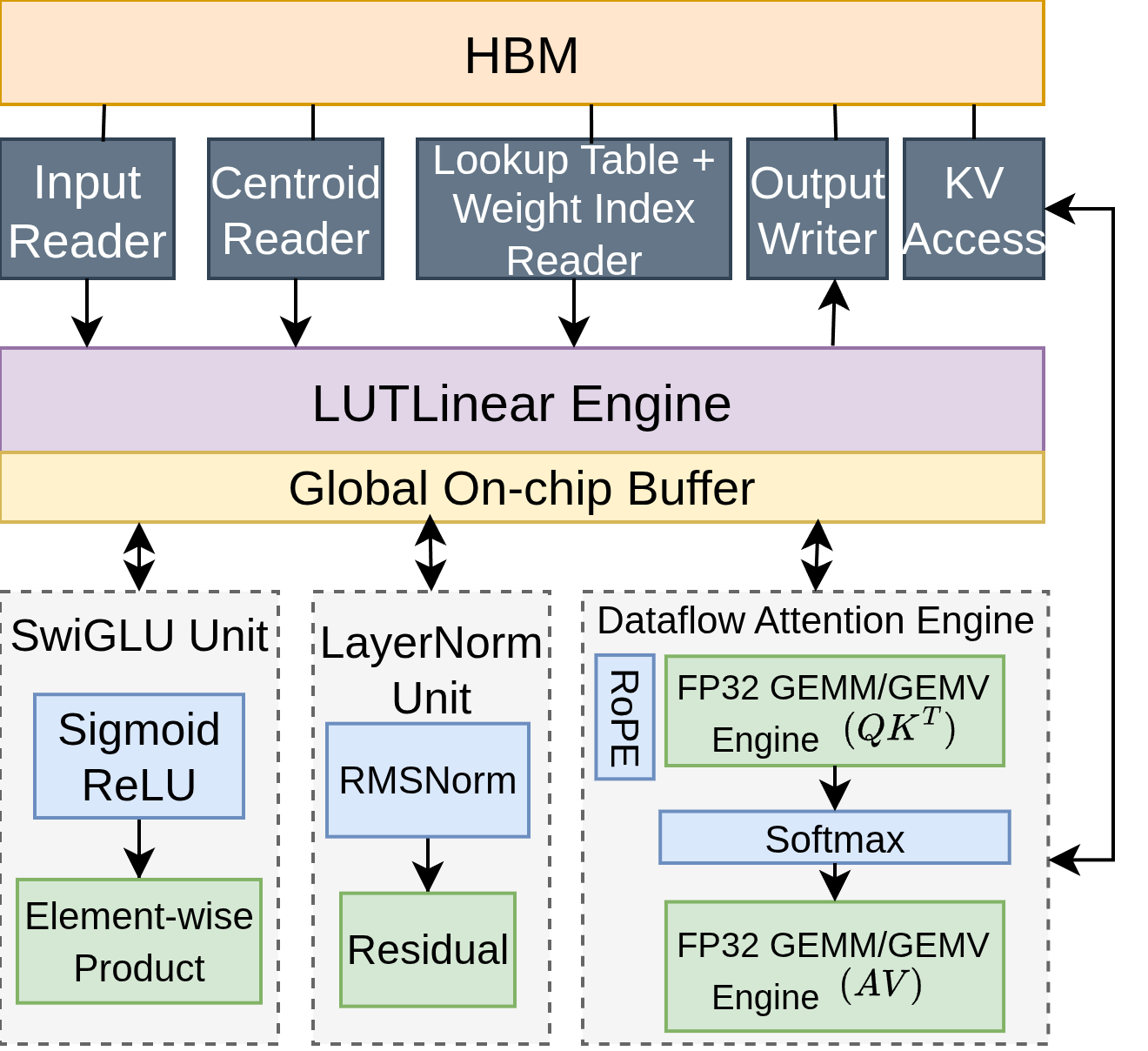}
    \caption{The overall architecture of LUT-LLM, including a LUTLinear engine with global buffer, a dataflow attention engine, and special functions (SwiGLU, LayerNorm) with pipelined operations. }
    \label{fig:lutlm}
\end{figure}

\begin{figure*}[ht]
    \centering
    \includegraphics[width=\linewidth]{  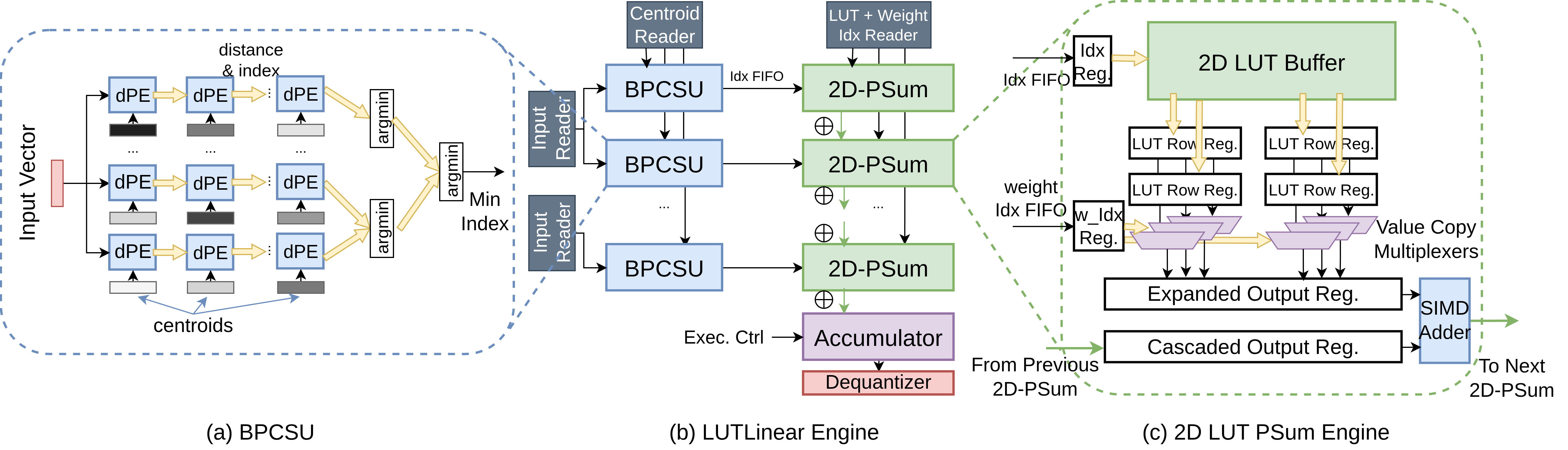}
    \caption{The LUTLinear Engine (b) performs all linear projections with bandwidth-aware parallel centroid search units (BPCSU) and 2D lookup table prefix-sum (2D-PSum) engines. The accumulator and dequantizer is used to aggregate and convert results to FP32. (a) BPCSU uses distance PEs (dPE) and a reduction tree to find the nearest centroids, with parallelism determined by the allocated memory bandwidth. (c) 2D-PSum retrieves centroid indices, gathers rows from the lookup tables, expands them with multiplexers, and performs SIMD addition before cascading to the next engine.}
    \label{fig:lutlinear}
\end{figure*}

The LUTLinear engine (Figure \ref{fig:lutlinear}(b)) consists of two primary components: bandwidth-aware parallel centroid search units (\textbf{BPCSU}) and a 2D lookup table prefix-sum engine (\textbf{2D-PSum}). BPCSU performs a nearest-neighbor search to identify the closest centroid for the given input vectors, while the 2D-PSum conducts a table lookup and accumulation to realize matrix multiplication. Each BPCSU is paired with a 2D-PSum, which receives the indices of the closest centroids. The 2D-PSum are connected in a serial chain to cascade and accumulate partial sums. Multiple BPCSUs and 2D-PSum enable parallel processing with multiple codebooks along the hidden dimension. The accumulated results are stored in the accumulator along with the output buffer. Before distributing outputs to other modules, the dequantizer converts each value in the output buffer to FP32 with the specified per-tensor scaling and shifting factors.

\subsection{Bandwidth-aware Parallel Centroid Search}

As shown in Figure \ref{fig:vq2d}, during inference, the first step is finding the closest centroids for the incoming input vectors. For hardware efficiency, the distance metric that is often used is the Chebyshev distance \cite{cantrell2000modern}, which calculates the distance between two vectors $\vec{a}$ and $\vec{b}$ by $L_\infty = \max_i (|a_i - b_i|)$, where $a_i$ and $b_i$ are elements in $\vec{a}$ and $\vec{b}$. To facilitate distance computation and comparison, previous work \cite{li2025lut, wei2025t} instantiated distance PEs (dPEs) for each centroid in the codebook. Since codebooks are varied across hidden dimensions and layers, every dPE needs to load new centroids. dPEs are connected in a single-chain of pipeline, where the last dPE in the chain produces the index of the closest centroid. Although having low overhead when a long sequence of input vectors is pipelined, this design incurs high setup latency if the sequence is short, e.g., in the decode stage. This latency is often higher than the off-chip memory access latency of lookup tables for LLMs. One potential alternative is to launch all dPEs to compute distance with all centroids in parallel, and apply a binary reduction tree to find the closest centroid. However, this significantly increases fanout and introduces $c_a-1$ comparators, which escalate routing difficulty, negatively impact timing, and consume more power on driving resources.

\begin{figure}[ht]
    \centering
    \includegraphics[width=0.95\linewidth]{  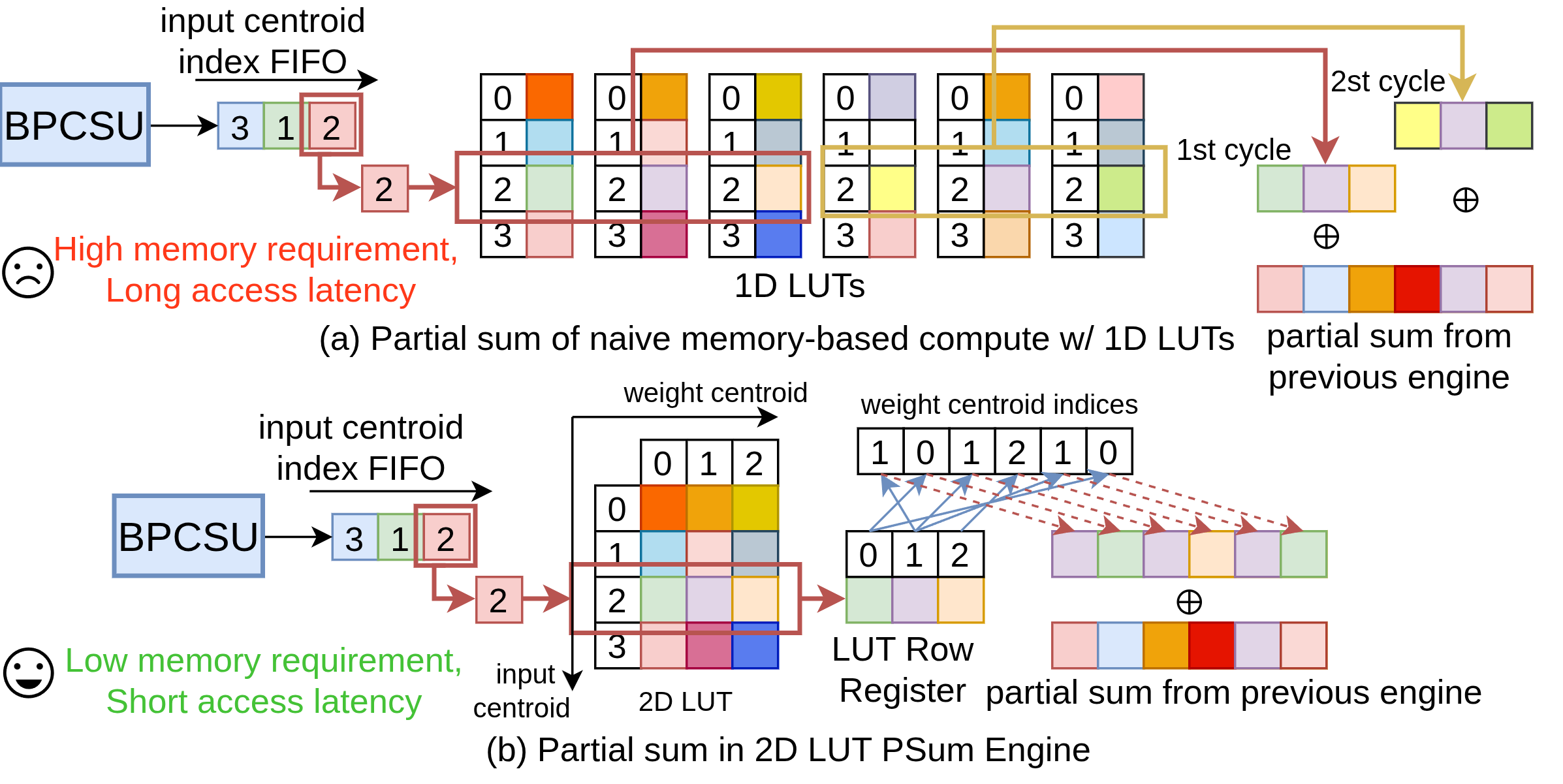}
    \caption{(a) Partial sum with 1D lookup tables requires more memory capacity to store the tables and longer latency to access data for accumulation. (b) Partial sum in 2D-PSum needs smaller memory and fewer memory ports to have high parallelism in accumulation.}
    \label{fig:2dlut_sample}
\end{figure}

In LUT-LLM, we propose the bandwidth-aware parallel centroid search unit (BPCSU), where we design the structure of data communication among dPEs to match the bandwidth of data loading in the 2D-PSum. We found that a complete reduction tree is often an overkill: we can support longer pipeline depth while the latency can still completely overlap with the off-chip memory access time. As seen in Figure \ref{fig:lutlinear}(a), BPCSU arranges dPEs into multiple pipeline chains. The input vector is broadcast to all chains for parallel distance computations. Each pipeline chain produces the minimum distance and index locally, which will be collected and compared through a small reduction tree. Consequently, BPCSUs can have a lower pipeline setup latency than data loading, with fewer resources instantiated for comparators than complete binary reduction trees. To maximize performance by overlapping centroid search and data loading, we have the following formulation: given $M$ as the output size, $G$ as the weight VQ group size, $c_w, c_a$ are the weight and activation codebook sizes, and $C$ is the memory bandwidth per BPCSU (bits/cycle), the length of each dPE pipeline chain $l$ is
\begin{equation}
    \arg \max_l(\frac{8c_ac_wM}{GC} + \log(c_w)\frac{M}{C}) \ge 32\frac{c_a}{C}+ l + \log(\frac{c_a}{l}))
\end{equation}
where lookup tables are in INT8 and centroids are in FP32. $\log(c_a/l)$ is the depth of the reduction tree. We evaluate the maximum $l$ that satisfies this condition for every linear projection and select the minimum over them. For example, for configurations of Qwen 3 1.7B, $l = 16$ for $c_a = 64$ and $c_w = 16$, indicating the instantiation of a $16\times 4$ array of dPEs with a 2-level reduction tree for each BPCSU.

\subsection{Efficient 2D Table lookup Based Prefix-Sum}

After retrieving the closest centroid index, LUT-LLM will perform table lookups and accumulations to compute matrix multiplications. 
Naive memory-based computation accesses 1D lookup tables for every element in the output dimension. With a limited number of memory ports (Figure \ref{fig:2dlut_sample}(a)), the partial sum takes multiple cycles. To execute table lookup efficiently, we construct 2D lookup tables for the dot-product outputs and perform lookups with a dedicated engine. Element $(i, j)$ in each table is the dot product between centroid $i$ in activation codebook and centroid $j$ in weight codebook. Illustrated in Figure \ref{fig:2dlut_sample}(b), given the incoming input centroid index received from BPCSUs, the 2D-PSum first accesses the row of 2D lookup tables with the corresponding index, and extracts to a LUT row register. Meanwhile, the centroid indices of the weight vectors are prefetched and stored in a register of length-$G$, where $G$ is the quantization group size. Then, the dot product result of each weight vector is retrieved and expanded by looking up the LUT row register using the weight centroid index. In the example, the engine will copy the value at position 1 of the LUT row register to position 0 in the output register for the first index in the weight centroid index register. Both the extraction of LUT row registers and data copying for expanded output registers finish in a single cycle with memory partitioning and pipelined 2D table lookup.

Figure \ref{fig:lutlinear}(c) illustrates the hardware architecture of the 2D-PSum. The 2D LUT buffer, which stores collections of lookup tables for each codebook, is instantiated utilizing a mixture of BRAM, URAM, and LUTRAM to balance memory resources on the target FPGA. Once the LUT row registers are extracted, their contents are forwarded to value copy multiplexers, which select entries according to the weight centroid index registers. To alleviate excessive fanout from the LUT row registers to the multiplexers, the registers are duplicated so that each copy drives only a subset of multiplexers. The expanded outputs are then accumulated with the results cascaded from the preceding 2D-PSum through a SIMD adder.

\subsection{Spatial-Temporal Hybrid Design}

\begin{figure}[ht]
    \centering
    \includegraphics[width=\linewidth]{  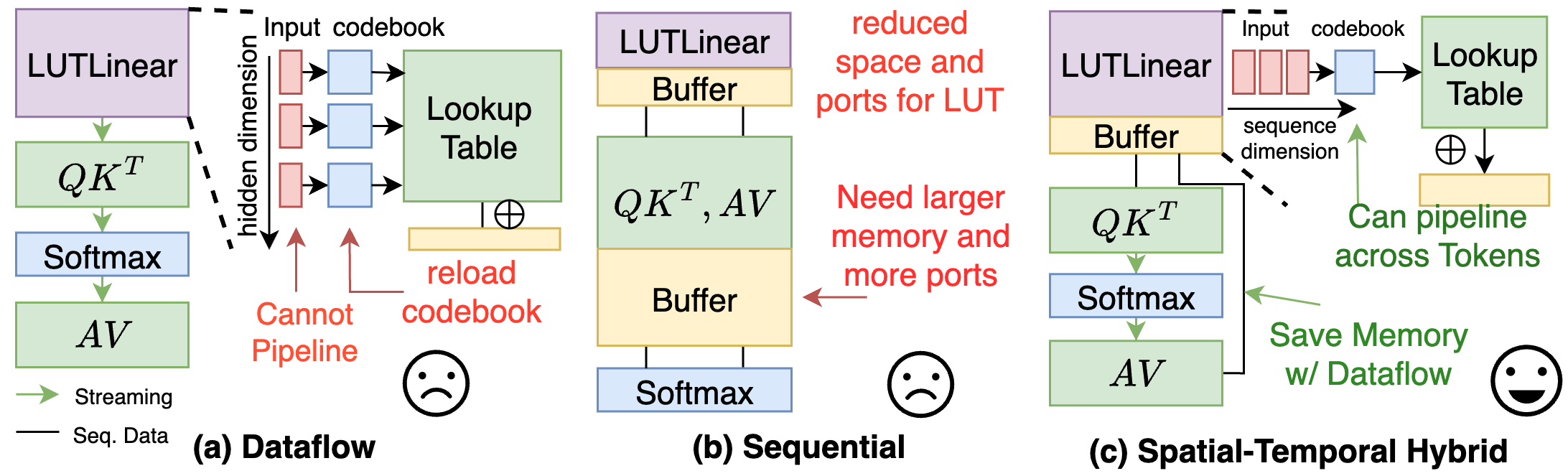}
    \caption{(a) Dataflow execution: centroid search cannot be pipelined across tokens, and repeated codebook reloading induces pipeline stalls. (b) Sequential execution: excessive memory resources are allocated to the attention block, which degrades linear projection throughput. (c) Hybrid execution: enables full token-level pipelining while minimizing memory footprint, allowing parallel table lookup and preserving high projection throughput.}
    \label{fig:spatial_temporal_hybrid}
\end{figure}

To integrate the LUTLinear engine into a complete language model accelerator with high efficiency, it is essential to carefully orchestrate inter-operation communication. Conventional language model accelerators typically adopt either pure dataflow \cite{chen2024allo, ye2025streamtensor} or sequential \cite{zeng2024flightllm, hong2022dfx} execution strategies. While both exhibit strong performance for arithmetic-intensive operations, they are suboptimal when the LUTLinear engine is introduced. For example, in a linear projection followed by attention computations, pure dataflow execution (Figure \ref{fig:spatial_temporal_hybrid}(a)) instantiates separate hardware modules for each operation and streams partial outputs through FIFOs. However, in the case of LUTLinear, generating partial outputs for streaming requires iterating over input vectors along the hidden dimension. This introduces two inefficiencies: (1) the same codebook must be reloaded multiple times, and (2) the centroid search cannot be pipelined, as the codebooks vary across the hidden dimension. These issues result in a significant stall latency in the linear layer. On the other hand, applying a purely sequential execution strategy (Figure \ref{fig:spatial_temporal_hybrid}(b)) requires the accelerator to allocate large buffers with high memory partitioning to compute and store attention scores. Consequently, the available memory capacity and port bandwidth for table lookups in LUTLinear are reduced, constraining the decoding and short-context prefilling speed since they are memory-bound. 

Inspired by previous accelerators \cite{he2025intar, zhuang2024ssr}, we design LUT-LLM to employ a spatial-temporal hybrid execution strategy (Figure \ref{fig:spatial_temporal_hybrid}(c)). The LUTLinear engine executes sequentially across different operations, while its outputs are delivered to subsequent modules, i.e., attention, SwiGLU, and LayerNorm, via dataflow execution. The data transfer logic is coarse-grain reconfigured similar to \cite{he2025intar} by execution control signal to the accmulator (Figure \ref{fig:lutlinear}(b)). This approach enables the LUTLinear engine to iterate over input vectors along the sequence dimension, sustaining a perfectly pipelined centroid search without codebook reloads. At the same time, the memory requirements for attention remain the same as dataflow execution, saving 14\% on-chip resources for high-throughput table lookups used in LUTLinear than sequential execution.

\subsection{Other Design Features}

\noindent \textbf{Efficient Attention for both Prefill and Decode Stages.} Since LUT-LLM performs the prefill and decode stage on the same design, ensuring high efficiency for both stages simultaneously can be challenging. The prefill stage can parallelize computations along the sequence for $Q$, $K$, and $V$, while the decode stage only has a single token $Q$ vector involved. In LUT-LLM, we parallelize computations only along the hidden dimension and sequence dimension for $K$ and $V$. This ensures full utilization of every compute element in both stages.

\noindent \textbf{KV Cache Prefetch and Write-out Orchestrations.} During the prefill stage, the KV cache is streamed to the HBM controller while delivering to the dataflow attention engine. During the decode stage, previously dumped KV cache is prefetched from HBM when the first linear projection is running. The KV embeddings of the new token will then be written to HBM at the end of the KV cache. 

\section{Experiment}

\subsection{Experiment Setup}

\noindent \textbf{Models, Datasets, and Algorithm.} Our overall scheme (following the notation in Table \ref{tab:symbols}) is: $G = 512$, $v = 2$, $c_w = 16$, $c_a = 64$, and INT8 quantized lookup tables. Based on Qwen 3 1.7B model \cite{yang2025qwen3}, we first calibrate for activation quantization. We applied per-tensor zero-point quantization \cite{jacob2018quantization} on the lookup tables, which quantizes the FP32 lookup tables $X$ into quantized matrix $X_{INT8}$ by
\begin{equation}
\begin{split}
    s = (\max(X)-\min(X))/256, z = - \min(X) / s \\
    X_{INT8} = \min(256, \max(0, sX+z))
\end{split}
\end{equation}
We optimize the two-stage training scheme of LUT-DLA \cite{li2025lut}. In the first stage, a fine-grained, layer-wise initialization of K-means is applied to obtain the activation centroids. This improves the overall training stability. Additionally, we design algorithmic optimizations for the forward and backward passes. In the forward pass, we construct a new lookup table gathering reduce kernel, which cuts memory usage with runtime lookup table reconstructions and accelerates the training process by fusing table lookups and accumulations. In the backward pass, we fuse the kernels that compute the gradient of weights and centroids. These updates allow us to train models beyond 1B parameters. Quantization-aware training is supported with the Straight-Through Estimator (STE) \cite{yin2019understanding} with adjustable gradients. With these optimizations, the total training time is reduced from over 1,000 A100 GPU hours to approximately 10 A100 GPU hours, representing an efficiency improvement over LUT-DLA. 

After training for activation quantization, we reconstruct the weights from the trained lookup tables, then apply GPTVQ \cite{van2024gptvq}. Then, we pre-compute the new lookup tables by computing the dot products between the vectors in activation codebooks and weight codebooks. 

For evaluation, we utilize the GLUE benchmark \cite{wang2018glue}, SQuAD v2 \cite{rajpurkar2016squad}, and MMLU-Pro \cite{wang2024mmlu} for question-answering dataset. Benchmarks are evaluated based on accuracy and F1 score. We compare with SmoothQuant \cite{xiao2023smoothquant} and SpinQuant \cite{liu2024spinquant} in W8A8 to represent scalar quantization schemes.

\noindent \textbf{Metrics.} We employ latency (end-to-end and decoding) and energy efficiency in tokens per Joule as metrics. We convert latency to relative speedup for better visualization. 

\noindent \textbf{Target Devices and Tools.} LUT-LLM is prototyped on AMD Alveo V80 FPGA \cite{amd2024alveov80}. To design the circuit, we use Xilinx \textbf{Vitis HLS} 2024.2 with TAPA framework \cite{guo2023tapa} and utilize RapidStream \cite{guo2022rapidstream, guo2021autobridge} for coarse-grain floorplanning to reduce routing congestion and improve timing. The design is implemented with Vivado 2024.2 with our customized block design to integrate the HLS-based RTL block and generate the bitstream. Resource utilizations is in Table \ref{tab:resource}.

\noindent \textbf{GPU Benchmarking.} We select AMD Instinct MI210 \cite{amd2025mi210} and NVIDIA A100 \cite{nvidia2021a100} as representatives. Table \ref{tab:hardware_config} compares V80 with the GPUs. The peak TOPs for V80 is scaled based on the achieved frequency. We use Huggingface Transformers for a naive GPU baseline and vLLM \cite{kwon2023efficient} with GPTQ \cite{frantar2022gptq} INT8 quantization as an optimized baseline for benchmarking. A100 has GPTQ Marlin kernels \cite{frantar2025marlin} integrated for efficient INT8 and INT4 inference. MI210 does not support this optimization. We utilize pyNVML \cite{nvidia-ml-py} and pyrsmi to monitor GPU power.

\begin{table}[t] 
\setlength{\tabcolsep}{4pt}
\centering
\caption{Hardware Configuration of FPGA and GPUs}
\label{tab:hardware_config}
\resizebox{0.75\linewidth}{!}{%
\begin{tabular}{lccc}
\toprule
 & \textbf{AMD V80} & \textbf{AMD MI210} & \textbf{NVIDIA A100} \\
\midrule
Frequency & 250 MHz & 1700 MHz & 1065 MHz \\
Max Bandwidth & 819 GB/s & 1.6 TB/s & 2.0 TB/s \\
Peak Power & 190 W & 300 W & 300 W \\
Peak INT8 TOPS & 25 & 181 & 624 \\
Peak BF16/FP16 TOPS & 5.3 & 181 & 312 \\
Peak FP32 TOPS & 5.3 & 45.3 & 19.5 \\
Process Node & 7 nm & 6 nm & 7 nm \\
\bottomrule
\end{tabular}%
}
\end{table}

\begin{table}[t]
\centering
\caption{Evaluation on GLUE, SQuAD v2, and MMLU-Pro datasets for Qwen 3 1.7B with different quantization schemes.}
\setlength{\tabcolsep}{4pt}
\resizebox{\linewidth}{!}{
\begin{tabular}{l|cccccc|cc}
\hline
\multirow{2}{*}{Config} & \multicolumn{6}{c}{\textbf{GLUE}} & \multicolumn{2}{c}{\textbf{QA}} \\
\cline{2-7} \cline{8-9}
& MNLI & MRPC & QNLI & QQP & RTE & SST-2 & SQuADv2 & MMLU-Pro \\
\hline
FP16 Baseline & 87.6 & 86.5 & 92.9 & 91.2 & 80.9 & 93.7 & 72.8 & 33.1 \\
RTN INT8 & 86.7 & 80.2 & 88.0 & 89.3 & 70.4 & 87.4 & 62.0 & 23.6 \\
SmoothQuant \cite{xiao2023smoothquant} & 87.0 & 85.3 & 91.7 & 89.6 & 79.1 & 91.2 & 71.3 & 31.7 \\
SpinQuant \cite{liu2024spinquant} & 87.3 & 83.3 & 91.8 & 89.5 & 80.2 & 91.5 & 72.0 & 28.0 \\
\textbf{LUT-LLM} & --- & --- & --- & --- & --- & --- & --- &  --- \\
+ Act. Quant. & 87.0 & 84.1 & 91.9 & 90.7 & 78.3 & 91.2 & 70.3 & 31.8 \\
+ INT8 LUT & 86.9 & 83.8 & 91.7 & 90.8 & 76.9 & 90.7 & 69.8 & 31.3 \\
+ Weight Quant. & 86.9 & 82.8 & 90.4 & 89.5 & 76.5 & 90.6 & 69.7 & 30.8 \\
\hline
\end{tabular}}
\label{tab:algo_results}
\end{table}



\noindent \textbf{SoTA FPGA Accelerator Baselines.} We select three representative SoTA works, Allo \cite{chen2024allo}, FlightLLM \cite{zeng2024flightllm}, and InTAR \cite{he2025intar}. They are arithmetic-based accelerators. Allo and InTAR apply the W4A8 scheme and FlightLLM utilizes 3.5-bit weights. FlightLLM employs aggressive sparsification and quantization on both weights and attention. This fundamentally alters the computation pattern and data loading size. Since these accelerators do not natively support the latest Qwen 3 1.7B model, we construct a simulator for Allo and InTAR and modify the simulator in FlightLLM's artifact to get their latency on V80, with less than a 2\% gap from data in the original papers.

\begin{figure}[t]
    \centering
    \includegraphics[width=0.9\linewidth]{  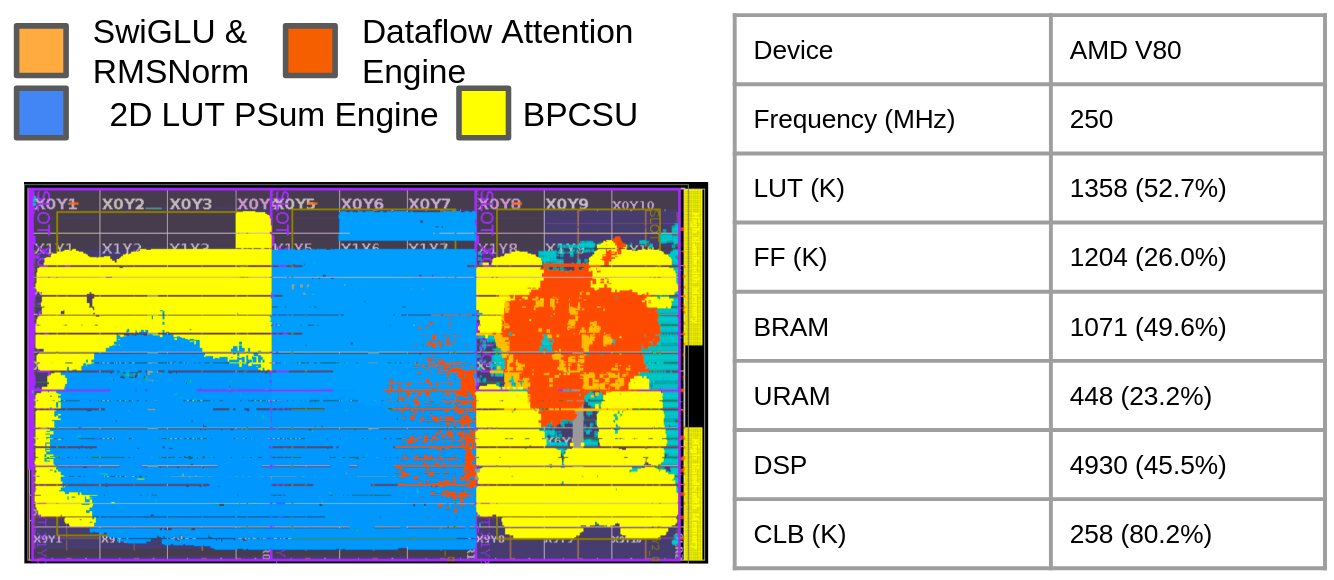}
    \caption{LUT-LLM implementation layout on AMD V80 FPGA, resource utilization, and clock frequency.}
    \label{fig:die_photo}
\end{figure}

\subsection{Algorithm Performance}

With the quantization scheme of LUT-LLM, the target model maintains a competitive performance compared with the FP16 baseline and consistently outperforms standard round-to-nearest (RTN) quantization to INT8 for activations. Table \ref{tab:algo_results} summarizes the model quality under different quantization schemes. Compared with SoTA scalar quantization technique, SpinQuant \cite{liu2024spinquant}, LUT-LLM has only $1.8\%$ drop in geomean accuracy. These results highlight the effectiveness of the quantization scheme and tuning methods used in LUT-LLM in preserving model quality and efficient deployment on FPGAs.

\subsection{Accelerator Performance}

\begin{figure}[ht]
    \centering
    \includegraphics[width=\linewidth]{  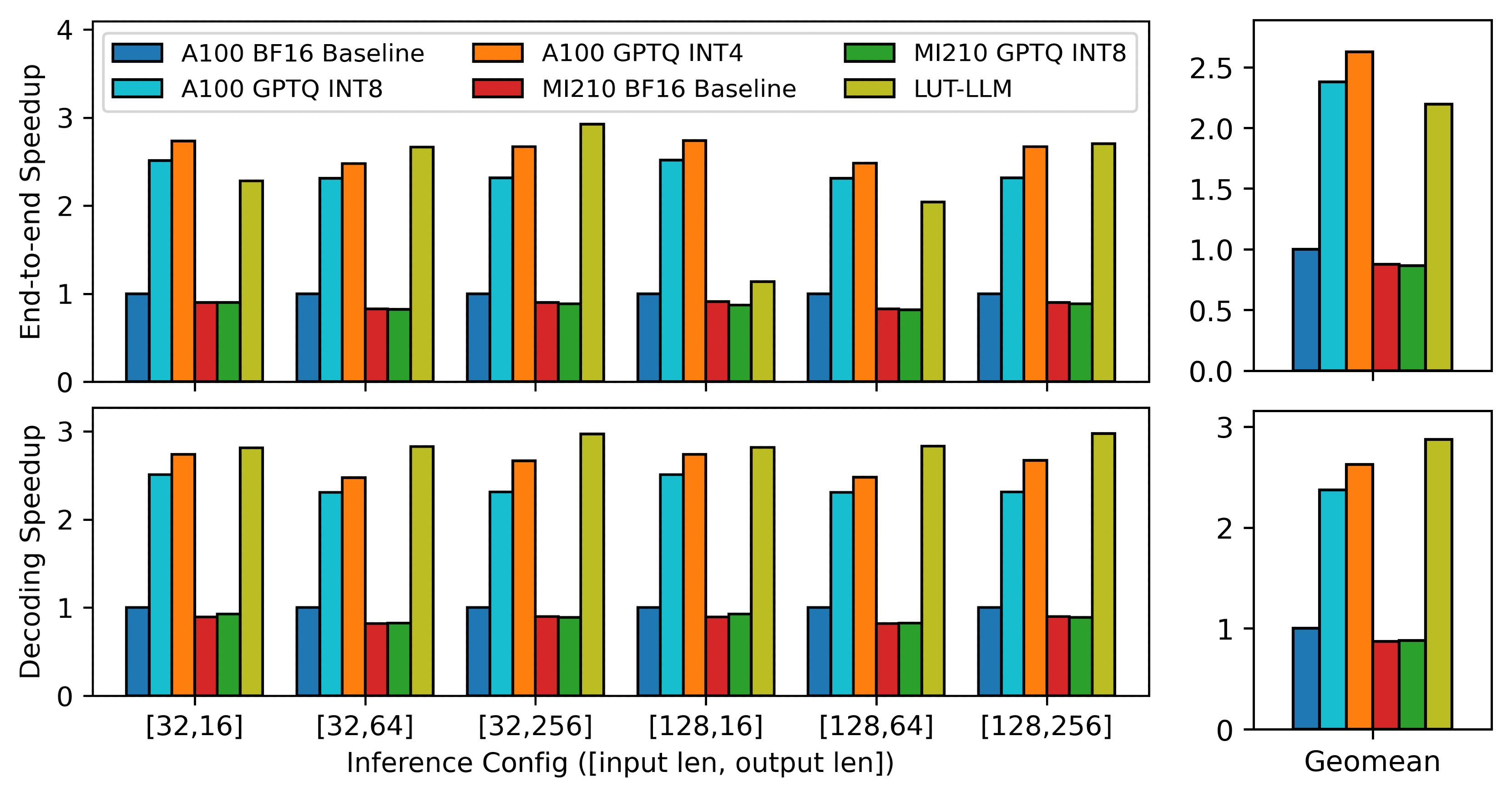}
    \caption{End-to-end and decode speedup of LUT-LLM and GPUs (both in 7nm process node) with BF16,  INT8, and INT4 precisions. }
    \label{fig:gpu_lat}
\end{figure}

\begin{figure}[ht]
    \centering
    \includegraphics[width=\linewidth]{  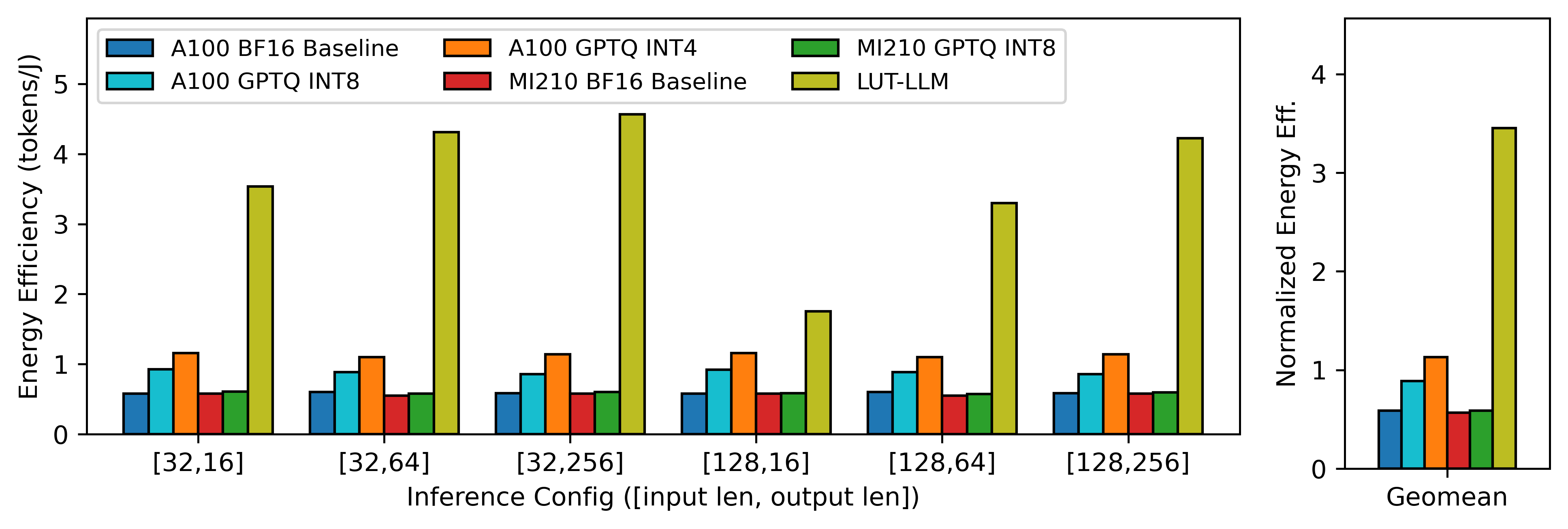}
    \caption{Energy efficiency (tokens per Joule) of LUT-LLM and GPUs with BF16, INT8, and INT4 precision.}
    \label{fig:energy_eff}
\end{figure}

\subsubsection{Resource Utilization}

Figure \ref{fig:die_photo} presents the floorplan of LUT-LLM on the AMD V80 FPGA. The design attains a clock frequency of 250 MHz. Although the utilization of all resource types remains at or below 50\%, the CLB utilization reaches 80\%. LUT-LLM employs 32 HBM channels for reading lookup tables and weight indices, along with 4 channels for input data and 4 channels for outputs and KV cache. Each channel on the V80 provides a 256-bit interface. The HBM clock is aligned with the kernel frequency to sustain a balanced read and write throughput. Consequently, the effective bandwidth utilized for loading lookup tables is $250 \times 256 \times 32 / 8 / 1024 = 250~\text{GB/s}$.

\subsubsection{Comparison with GPUs}

Figure \ref{fig:gpu_lat} reports the end-to-end latency and decoding speedup across all configurations. Leveraging quantized lookup tables, LUT-LLM achieves a $3.29\times$ lower geometric-mean latency than the MI210 with INT8 quantization. On the A100, LUT-LLM delivers a $1.46\times$ speedup over the FP16 baseline and $1.21\times$ over the INT8 baseline. Notably, even compared to A100 with INT4 quantization, which reduces model storage by half relative to LUT-LLM, our approach still attains a $1.10\times$ decoding speedup.

The limited speedup of INT4 on GPUs compared to INT8 can be attributed to two factors. First, vLLM employs the Marlin kernel for INT4 execution, which performs on-the-fly dequantization from INT4 to FP16 and subsequently utilizes FP16 tensor core instructions for multiply-accumulate operations. This dequantization overhead and reliance on higher-precision compute reduce the achievable throughput compared to native low-precision execution. Second, effective HBM bandwidth utilization on GPUs degrades for smaller models at lower precision. We observed that with a 1.7B model in INT4, the A100 attains only $0.6\times$ of the bandwidth utilization than with INT8. We acknowledge that this diminished scaling may not hold for larger models \cite{wang2025systematic}: when extrapolated to Qwen 3 32B, LUT-LLM achieves $1.2\times$ speedup over A100 with INT8, but only $0.8\times$ when deploying the INT4 model on A100.

 We then compare the energy efficiency, as shown in Figure \ref{fig:energy_eff}. While all GPUs benefit from quantization due to reduced off-chip memory traffic and specialized low-bit tensor core, LUT-LLM maintains a substantial advantage, achieving $6.6\times$ higher geomean energy efficiency than MI210. For A100, LUT-LLM remains $3.05 \sim 5.94\times$ more energy efficient in geomean. Moreover, the efficiency gap between LUT-LLM and GPUs widens as output length increases, underscoring the scalability of LUT-LLM for long output sequences.

\subsubsection{Comparison with SoTA FPGA Accelerators}

\begin{figure}[t]
    \centering
    \includegraphics[width=0.9\linewidth]{  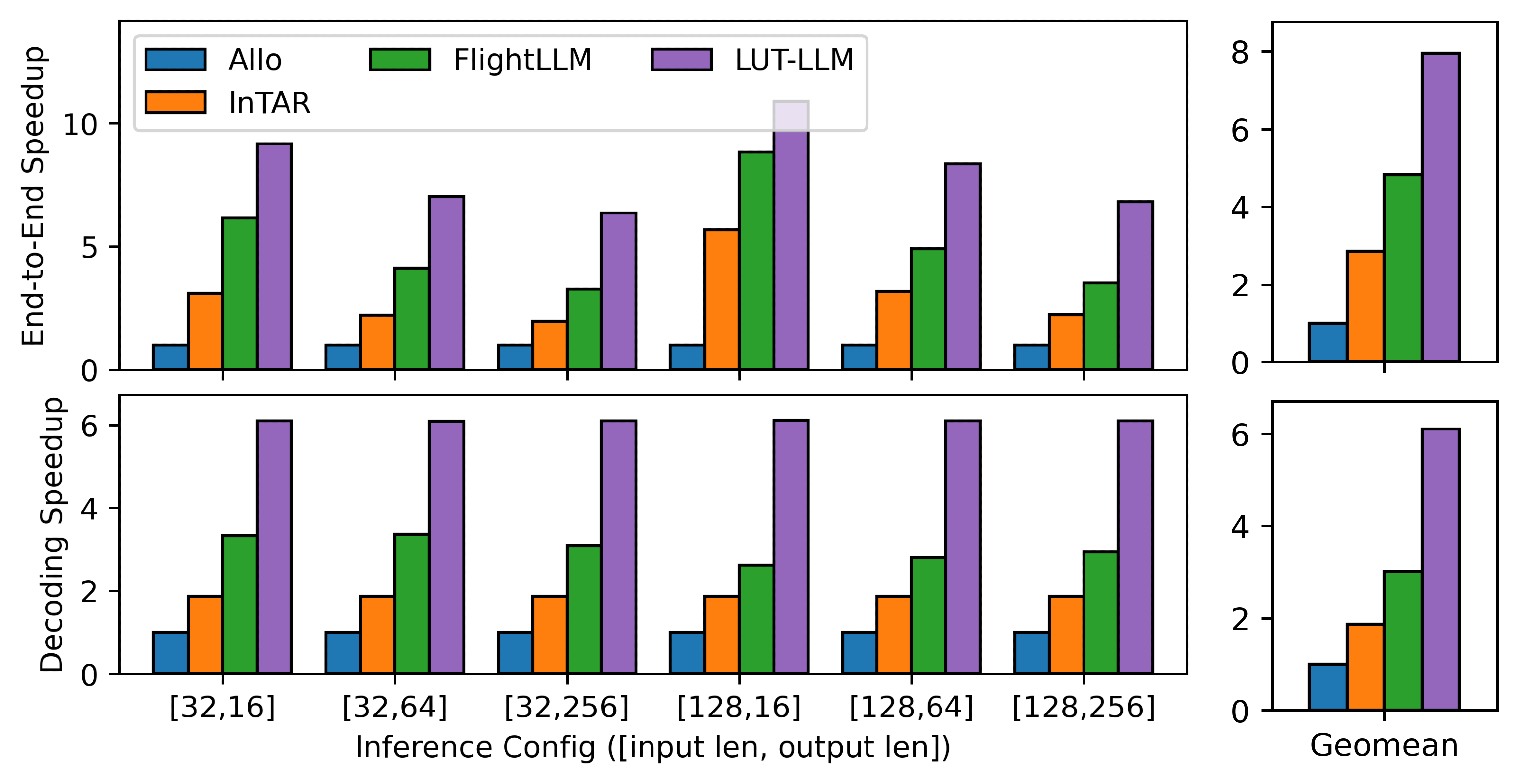}
    \caption{Speedup of LUT-LLM over InTAR, Allo, and FlightLLM.}
    \label{fig:sota_acc}
\end{figure}


Figure \ref{fig:sota_acc} compares the latency of LUT-LLM with InTAR, Allo, and FlightLLM. Overall, LUT-LLM is $5.6\times$ and $1.9\times$ faster than Allo and InTAR. The performance gain arises from improvements in both the prefill and decode stages. In the prefill stage, LUT-LLM leverages on-chip memory access as computation for higher peak TOPs than Allo and InTAR. In the decode stage, LUT-LLM performs computations more efficiently, fully utilizing the available memory bandwidth. Moreover, Allo and InTAR incur additional overhead from dequantization, which requires extra data loading and costly floating-point computations. Compared with FlightLLM, LUT-LLM is $1.6\times$ faster than FlightLLM end-to-end, with larger speedup as sequence length grows. 





\section{Conclusion and Outlooks}

This work presents LUT-LLM, the first FPGA accelerator for 1B+ LLM inference with memory-based computation. We develop a performance model and demonstrate that co-quantization is essential for speedup. LUT-LLM achieves $1.10\sim 3.29\times$ higher speed than GPUs. LUT-LLM can be mapped to ASICs by replacing FPGA-specific IPs. It is also promising for future memory-stacked designs \cite{chiang2025integration}.

\section*{Acknowledgement}

This work was supported in part by PRISM, one of the seven centers in the JUMP 2.0 program sponsored by SRC and DARPA. It is also supported by CDSC industrial partners and the AMD \footnote{J. Cong has a financial interest in AMD} HACC Program.

\section{Appendix: Performance Derivation} \label{sec:derivation}

\noindent \textbf{Weight Quantization.} Consider $G$-vectors quantization group with length-$v$ vectors and $c_w$ centroids per codebook. An $M\times D$ weight matrix is quantized to $4MDc_w/Gv$ bytes codebook in FP32 and $MD\log(c_w)/8v$ bytes weight centroid indices, which derives the latency of loading weights from off-chip as equation \ref{eq:w_mem}. Then, the accelerator will access the FP32 values in the codebooks based on the weight indices and copy them to the registers. To expand $S$ values, we need to access $S/v$ indices and $S / Gv$ codebooks, which requires $S\log(c)/v + 32S/Gv$ bits access per cycle. Consequently, we can expand $S = N_{p}b_{p}/(\log(c_w)/v+32/Gv)$ elements per cycle. Finally, the accelerator performs FP32 MAC operations and accumulates results to output buffer. The maximum number of operations per cycle is determined by both the compute units and memory ports available, which is $\min(N_{c}Op_{fp32}, N_{p}b_{p}/32)$. This concludes the latency of on-chip computation as shown in Equation \ref{eq:w_lat}.

\noindent \textbf{Activation Quantization.} Consider length-$v$ vectors with $c_a$ centroids per codebook. Since every $v$ channels in the hidden dimension and each output dimension require a separate lookup table, an $M \times D$ weight matrix is converted into $MD/v$ tables, each containing $c_a$ INT8 entries. The activation codebooks consist of $Dc_a/v$ vectors in FP32. Hence, the off-chip memory latency is given in equation~\ref{eq:a_mem}. Computation proceeds in two pipelined steps. First, the accelerator finds the closest centroid in the codebook for each input vector, requiring at least $\log(c)$ cycles via binary reduction. For $L$ tokens, this can be pipelined in $\log(c)+L-1$ cycles. The resulting indices are streamed for table lookup, where each index drives up to $M$ outputs but is limited by the available memory ports. Thus, table lookups for $L$ inputs take $ML/\min(M, N_{p}b_{p}/8)$ cycles. In parallel, the extracted values are accumulated in the output buffer, constrained by compute and memory resources. With $S$ parallel centroid searches across $D$, at least $Sc_av/Op_{fp32}$ units are needed, since each distance computation costs $v/Op_{fp32}$ and must be performed for all centroids. The remaining units perform $(N_{c}-Sc_av/Op_{fp32})Op_{int8}$ INT8 accumulations in parallel. Consequently, the latency of the table lookup per $S$-way search is as in equation~\ref{eq:a_tl}, and the total on-chip computation latency follows equation~\ref{eq:a_lat}.

\noindent \textbf{Activation-Weight Co-quantization.} Consider the settings the same as the weight and activation quantization. The accelerator needs to load $MD/Gv$ tables with $c_wc_a$ entries each, $MD\log(c_w)/8v$ bytes of weight indices, and $4Dc_a/v$ bytes of activation codebooks. We can then derive equation \ref{eq:2d_mem} and \ref{eq:2d_lat} by following the same analysis as weight and activation quantization.



%
\bibliographystyle{IEEEtran}
\bibliography{references}

\end{document}